\documentstyle{article}
\def\c{\,\cdot\,}
\def\w{\overline w}
\def\u{\overline u}
\def\au{\acute u}
\def\aw{\acute w}
\def\nau{\acute{\overline u}}
\def\naw{\acute{\overline w}}
\def\ac{\acute}
\def\ds{\displaystyle}
\begin{document}

\phantom{Draft}
\vspace{3cm}

\begin{center}
{\Large\bf
On a two dimensional system associated with the
complex of the solutions of the Tetrahedron equation.
}
\end{center}

\vspace{.5cm}

\begin{center}
S. M. Sergeev\\ \vspace{0.5cm}
Branch Institute for Nuclear Physics, Protvino 142284, Russia.\\
E-mail: sergeev\_ms@mx.ihep.su
\end{center}

\vspace{1cm}

{\small
\noindent
{\bf{Abstract}}:
A sort of two dimensional linear auxiliary problem
for the complex of 3D $R$ -- operators associated with
the Zamolodchikov -- Bazhanov -- Baxter statistical model
is proposed. This problem resembles
the problem of the local Yang -- Baxter equation
but does not coincide with it.
The formulation of the auxiliary problem
admits a notion of a ``fusion'', and usual
local Yang -- Baxter equation appears among other
results of this ``fusion''.
}

\bigskip

\newpage

\section{Introduction}

Well known shortcoming of the three dimensional integrability,
connected with the tetrahedron equation, is the
remarkable poverty of the models. Actually, there exists only one
nontrivial complex of solutions of the tetrahedron equation,
which has statistical mechanics, field theory
and ``classical'' (i. e.  functional) forms. In the statistical
mechanics, when the number of states is finite,
the model is called Zamolodchikov -- Bazhanov -- Baxter
model \cite{zam-solution,bb-first}. It is known almost everything
(except the most complicated aspects, concerning the
structure of eigenstates of the transfer matrices and
the correlation functions) about this model, namely:
the partition function \cite{baxter-pf,bb-first},
the vertex -- IRF correspondence \cite{mss-vertex,mss-psi},
affine Toda field theory \cite{kr-ats} and
field theory operator $R$ -- matrix \cite{fk-qd,sbm-qd},
and the transition from
the infinite number of the states to finite one
\cite{br-qd,sbm-qd},
and finally, the functional $R$ -- operator
as the sequence of the permutation relations for
the field theory $R$ -- matrix \cite{sbm-qd,ms-modified}.
Here one should mention that there exists an hierarchy
of the operator $R$ -- matrices which corresponds to
several partial specifications of the spectral parameters
of ZBB matrix weights. There must exist a reverse way,
obtaining of more complicate $R$-s in the hierarchy
in terms of simplest (some sort of the fusion).
Also one may mention the interpretation
of the projection of 3D operator $R$ -- matrices in terms
of a representation of some specific Drinfeld double
\cite{double}.

From the other hand side there are a lot of solutions of the
functional tetrahedron equation. Moreover, we've got infinitely many
such solutions \cite{korepanov-diss}
{\em versus} one specific given by
the  complex  mentioned above.

Recall that usually the functional tetrahedron solution appears
when one considers so -- called local Yang -- Baxter equation (LYBE)
\cite{maillet},
which is a proper generalization of the zero curvature condition
for two dimensional system. Namely, let $L_{i,j}(\vec x)$ be matrices of
weights of the Yang -- Baxter type, and $\vec x$ stands for their
formal parameters. Then LYBE
\begin{equation}\label{lybe}
L_{12}(\vec x)\c L_{13}(\vec y)\c L_{23}(\vec z)\;=\;
L_{23}(\ac{\vec z})\c L_{13}(\ac{\vec y})\c L_{12}(\ac{\vec x})\;,
\end{equation}
when it is nondegenerative with respect to
$\ac{\vec x},\ac{\vec y},\ac{\vec z}$,
gives the functional map from $\{\vec x,\vec y,\vec z\}$ to
$\{\ac{\vec x},\ac{\vec y},\ac{\vec z}\}$ by
\begin{equation}
\ds
\ac{\vec x}\;=\;\vec r_1(\vec x,\vec y,\vec z)\;,\;\;\;
\ac{\vec y}\;=\;\vec r_2(\vec x,\vec y,\vec z)\;,\;\;\;
\ac{\vec z}\;=\;\vec r_3(\vec x,\vec y,\vec z)\;,
\end{equation}
Usually the functional operator $R_{i,j,k}$, giving this map, is introduced:
\begin{equation}
R_{i,j,k}\c\phi(\vec x_i,\vec x_j,\vec x_k)\;=\;
\phi(\ac{\vec x}_i,\ac{\vec x}_j,\ac{\vec x}_k)\;,
\end{equation}
\begin{equation}
\ds
\ac{\vec x}_i=\vec r_1(\vec x_i,\vec x_j,\vec x_k),\;\;\;
\ac{\vec x}_j=\vec r_2(\vec x_i,\vec x_j,\vec x_k),\;\;\;
\ac{\vec x}_k=\vec r_3(\vec x_i,\vec x_j,\vec x_k)\;,
\end{equation}
where $\phi$ is an arbitrary function. Considering the quadrilateral
formed by six $L$-s
\begin{eqnarray}
&\ds
L_{12}(\vec x_1)\c L_{13}(\vec x_2)\c L_{23}(\vec x_3)\c
L_{14}(\vec x_4)\c L_{24}(\vec x_5)\c L_{34}(\vec x_6)\;=
&\nonumber\\
&\ds
=\;L_{34}(\vec y_6)\c L_{24}(\vec y_5)\c L_{14}(\vec y_4)\c
L_{23}(\vec y_3)\c L_{13}(\vec y_2)\c L_{12}(\vec y_1)\;,
&
\end{eqnarray}
one obtains rhs by two different ways, first starting from
$L_{12}L_{13}L_{24}$, second starting from $L_{23}L_{24}L_{34}$. This
gives the equivalence of two successive applications of $R_{i,j,k}$:
\begin{eqnarray}
&\ds \phi[\vec y_1,\vec y_2,\vec y_3,\vec y_4,\vec y_5,\vec y_6]\;=
&\nonumber\\
&\ds R_{123}\c
\Biggl(R_{145}\c
\Bigl( R_{246}\c
\bigl( R_{356}\c
\phi[\vec x_1,...,\vec x_6]\bigr)\Bigr)\Biggr)
\;=&\nonumber\\
&\ds=\;
R_{356}\c
\Biggl(R_{246}\c
\Bigl(R_{145}\c
\bigl(R_{123}\c
\phi[\vec x_1,...,\vec x_6]\bigr)\Bigr)\Biggr)
\;.&
\end{eqnarray}
This is the functional tetrahedron equation (FTE).
\footnote{
Here one should mention the successful attempt to
obtain 3D $R$ -- matrix directly from the similar consideration
of the Zamolodchikov -- type algebra for two dimensional
$L^\alpha_{i,j}$, where $\alpha$-s -- indices of 3D $R$ -- matrix
\cite{korepanov}. This $R$ -- matrix obtained appeared to be
some special case of the $R$ -- matrix for ZBB model.}

In the case when the local weights $L_{a,b}(\vec x)$ have the structure
of the ferro-electric type free fermion model's weights,
the irreducible part of eq. (\ref{lybe}) can be extracted
in the form of Korepanov's equation
\begin{equation}\label{korepanov}
X_{12}(\vec x)\c X_{13}(\vec y)\c X_{23}(\vec z)\;=\;
X_{23}(\ac{\vec z})\c X_{13}(\ac{\vec y})\c X_{12}(\ac{\vec x})\;,
\end{equation}
where `$\c$' means the matrix multiplication of
\begin{eqnarray}
&\ds X_{12}(\vec x)\;=\;
\left(\begin{array}{ccc}
a(\vec x)& b(\vec x)&  0\\
c(\vec x)& d(\vec x)&  0\\
0  & 0  &  1
\end{array}\right)\;,
&\nonumber\\
&\ds X_{13}(\vec x)\;=\;
\left(\begin{array}{ccc}
a(\vec x)& 0& b(\vec x)\\
  0& 1&   0\\
c(\vec x)& 0& d(\vec x)
\end{array}\right)\;,
&\nonumber\\
&\ds X_{23}(\vec x)\;=\;
\left(\begin{array}{ccc}
 1&   0&   0\\
 0& a(\vec x)& b(\vec x)\\
 0& c(\vec x)& d(\vec x)
\end{array}\right)\;,&
\end{eqnarray}
$a(\vec x),...,d(\vec x)$ are in general some {\em matrix} functions.
Eq. (\ref{korepanov}) and its connection with the functional
tetrahedron equation was investigated in \cite{korepanov-diss}
in most general case. There was proven that there exists
a wide class of the solutions of FTE associated with eq.
(\ref{korepanov}).
A subset of the solutions of FTE was described in
\cite{rmk-lybe}, \cite{ks-fun} and resently in
\cite{oneparam}. These are the cases
when $a,b,c,d$ are some numeric functions of single variable
$x$. Nevertheless, the solution of FTE associated with the ZBB
complex is still not described in terms of LYBE (moreover, we suspect,
this is impossible at least in  terms of the Korepanov's equation
(\ref{korepanov})).

So, {\bf what we wished to say by this long introduction.}
From one hand side a plenty of the solutions of FTE exist,
these solutions are associated with LYBE (LYBE makes FTE obvious),
but we have no skill to quantize them. From the other hand side,
only one model is the whole complex, i. e.
it exists in the statistical mechanics form,
field theory form and a functional transformation form, moreover,
there exists a well defined way to obtain the statistical
mechanics and  the field theory from the functional form,
but we do not know a sort of a linear problem
(like LYBE) for this.

In this paper we try to give the answer
to all these questions. We suggest some formulation of two dimensional
system associated with 2D lattice,
which is not Yang -- Baxter type system (i. e. we do not
interpret the vertices as matrices of weights or anything
equivalent), but nevertheless the usual graphic pictures remain
as well as a sort of zero curvature condition, FTE and so on.
Our formulation resembles the electric networks,
when the star -- triangle
equivalence gives the electric network
solution of FTE (well described
in terms of Korepanov's equation \cite{rmk-lybe,oneparam})
but originally this equivalence is not LYBE at all,
it is just a resoldering of the resistances.
What we suggest,
it is just a proper generalization of Kirchhoff's rules in terms
of arbitrary 2D networks formed by rectangular vertices.
Such formulation is not equivalent to the usual LYBE-s in general.
The functional transformation corresponding to
ZBB complex is a partial case of our functional transformation.
There is a sort of fusion in the system proposed,
and the ferro-electric case is just a
subspace of a partial case of fused system.

\section{Formulation of the system and $R$ -- operator}

\begin{figure}
\begin{center}
\setlength{\unitlength}{0.25mm} 
\thicklines
\begin{picture}(300,200)
\put(50,0)
{  
   \begin{picture}(200,200)
   \put(75,0){\vector(0,1){200}}
   \put(37.5,25){\vector(3,2){150}}
   \put(187.5,75){\vector(-3,2){150}}
   \put(75,50){\circle*{5}}
   \put(75,150){\circle*{5}}
   \put(150,100){\circle*{5}}
   \put(55,175){$J'_1$}
   \put(85,155){$\Phi'_1$}
   \put(55,135){$\Phi_1$}
   \put(80,120){$J_1$}
   \put(145,120){$J'_2$}
   \put(145,70){$J_2$}
   \put(175,95){$\Phi'_2$}
   \put(120,95){$\Phi_2$}
   \put(50,55){$\Phi_3$}
   \put(55,20){$J_3$}
   \put(80,75){$J'_3$}
   \put(80,35){$\Phi'_3$}
   \end{picture}
} 
\end{picture}
\end{center}
\caption{ The left -- hand side with the currents attributed to
each vertex.}
\label{fig-lhs}
\end{figure}
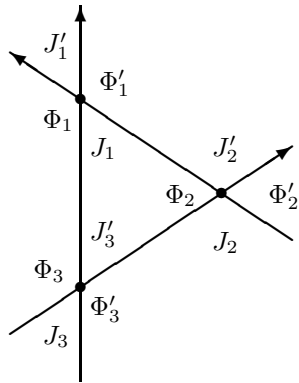

Consider $2D$ graphs formed by the usual oriented rectangular vertices.
Each vertex has four adjacent fields. Assign to these fields
some variables, {\em currents},
for the vertex shown in fig. \ref{fig-vertex} denote them
$J,J';\Phi,\Phi'$.
Suppose these current are additive for any field of the $2D$
network, and for any closed field let
the sum of the currents belonging to this field
and corresponding to the surrounding
vertices be zero (it is supposed that all the currents assigned to
any vertex flow into this vertex) -- see fig. \ref{fig-lhs} and
considerations for it for an example.
Suppose also the currents assigned to the vertex $V_k$ obey some condition,
$V_k(J_k,J'_k,\Phi_k,\Phi'_k)=0$ (obviously,
this condition must be a set of linear relations homogeneous
with respect to the currents).
Thus, for any \underline{open} network
there appears some condition for \underline{outer} currents.
Two networks are equivalent if their such conditions coincide.
For example, the evolution of 2D network is given by the
condition of the equivalence of the left hand side type
and right hand side type graphs, as it is drawn in fig. (\ref{fig-ybe}).
Solving such equivalence condition with respect to the parameters
of $V_k'$ in rhs, we obtain a solution of FTE.

\noindent
Now fix the form of $V(J,J',\Phi,\Phi')$ (a sort of Ohm's law):
\begin{eqnarray}\label{ohm}
&\ds
\Phi \;=\;i\,\w\,\cdot\,J\;,\;\;\;\;
\Phi'\;=\;-i\,u\,\cdot\,J\;,&\nonumber\\
&\ds
\Phi\;=\;-i\,\u\,\cdot\,J'\;,\;\;\;\;
\Phi'\;=\;i\,w\,\cdot\,J'\;,
\end{eqnarray}
with the primitive ``zero -- curvature'' condition:
\begin{equation}
u^{-1}\c w^{}\;=\;\w^{\,-1}\c \u^{}\;.
\end{equation}
Surely, invertible elements $u,w,\u,\w$ are treated as
the parameters of the
vertex $V$. Note also that this is not unique choice of
$V(J,J',\Phi,\Phi')$, we may for example suppose two of the currents
to be independent. Such cases we do not investigate here.

\begin{figure}
\begin{center}
\setlength{\unitlength}{0.25mm} 
\thicklines
\begin{picture}(350,150)
\put(100,0)
{  
   \begin{picture}(150,150)
   \put(15,75){\vector(1,0){120}}
   \put(75,15){\vector(0,1){120}}
   \put(75,75){\circle*{5}}
   \put(100,100){$\ds J'$}
   \put(100,40){$\ds \Phi'$}
   \put(40,100){$\ds \Phi$}
   \put(40,40){$\ds J$}
   \end{picture}
} 
\end{picture}
\end{center}
\caption{ Vertex $V$ with the surrounding attributed
currents $J,J',\Phi,\Phi'$.}
\label{fig-vertex}
\end{figure}
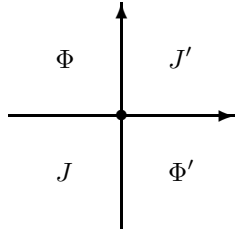

\begin{figure}
\begin{center}
\setlength{\unitlength}{0.25mm} 
\thicklines
\begin{picture}(450,200)
\put(00,0)
{  
   \begin{picture}(200,200)
   \put(75,0){\vector(0,1){200}}
   \put(37.5,25){\vector(3,2){150}}
   \put(187.5,75){\vector(-3,2){150}}
   \put(75,50){\circle*{5}}
   \put(75,150){\circle*{5}}
   \put(150,100){\circle*{5}}
   \put(85,155){$V_{12}$}
   \put(85,30){$V_{23}$}
   \put(140,115){$V_{13}$}
   \put(50,180){$x$}
   \put(25,95){$y'$}
   \put(50,15){$z$}
   \put(125,155){$z'$}
   \put(125,35){$x'$}
   \put(185,95){$y$}
   \end{picture}
} 
\put(220,95){$\ds =$}
\put(250,0)
{  
   \begin{picture}(200,200)
   \put(125,0){\vector(0,1){200}}
   \put(12.5,75){\vector(3,2){150}}
   \put(162.5,25){\vector(-3,2){150}}
   \put(50,100){\circle*{5}}
   \put(125,50){\circle*{5}}
   \put(125,150){\circle*{5}}
   \put(40,115){$\ac V_{13}$}
   \put(135,55){$\ac V_{12}$}
   \put(135,130){$\ac V_{23}$}
   \put(75,155){$x$}
   \put(5,95){$y'$}
   \put(75,40){$z$}
   \put(140,180){$z'$}
   \put(140,15){$x'$}
   \put(170,95){$y$}
   \end{picture}
} 
\end{picture}
\end{center}
\caption{ Baxter -- type equivalence of two graphs formed by the
vertices $V_{ij}$ in lhs and by $\ac V_{ij}$ in rhs and
outer currents $x,y,z,x',y',z'$.}
\label{fig-ybe}
\end{figure}
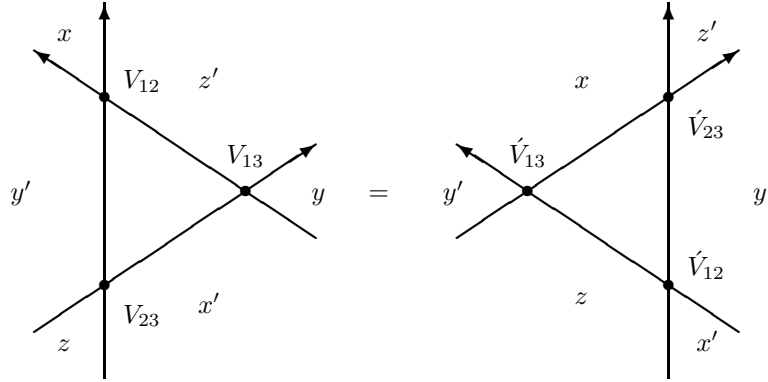

Consider now the Baxter -- type correspondence. Let the outer
currents be $x,y,z,x',y',z'$ as it is shown
in fig. \ref{fig-ybe}. Left -- hand side is drawn in fig. (\ref{fig-lhs}).
There $J_k,J'_k,\Phi_k,\Phi'_k$ are connected by (\ref{ohm}).
Conservation of the currents for closed field is
\begin{equation}
J_1+\Phi_2+J'_3\;=\;0\;,
\end{equation}
and given outer currents are
\begin{eqnarray}
&\ds J'_1\;=\;x\,,\;\;\; \Phi'_2\;=\; y\,,\;\;\;
J_3\;=\;z\,,&\nonumber\\
&\ds J_2+\Phi'_3\;=\;x'\,,\;\;\;
\Phi_1+\Phi_3\;=\; y'\,,\;\;\;
\Phi'_1+J'_2\;=\;z'\,.&
\end{eqnarray}
These with (\ref{ohm}) give the following system:
\begin{eqnarray}
&\ds u_1^{-1}\c w^{}_1\c x \;+\;
\u^{}_2\c w_2^{-1}\c y \;+\;
w_3^{-1}\c u^{}_3\c z\;=\;0\;,&\nonumber\\
&\ds
x'\;=\;i\,u_2^{-1}\c y \;-\; i\,u^{}_3\c z\;,
&\nonumber\\
&\ds
y'\;=\;-i\,\u_1^{}\c x \;+\; i\,\w_3^{}\c z\;,
&\nonumber\\
&\ds
z'\;=\;i\, w_1^{}\c x \;-\; i\, w_2^{-1}\c y\;.&\label{lhs}
\end{eqnarray}
Thus two of the currents are independent, and the rest four one can
express in terms of the independent.

Analogous consideration for the right -- hand side with
primed parameters of the vertices dives
\begin{eqnarray}
&\ds
\aw_1^{-1}\c \au_1^{}\c x' \;+\;
\aw_2^{}\c \nau_2^{-1}\c y' \;+\;
\au_3^{-1}\c \aw_3^{}\c z'\;=\;0\;,&\nonumber\\
&\ds
x\;=\;i\,\nau_2^{-1}\c y' \;-\; i\,\nau_3^{}\c z'\;,
&\nonumber\\
&\ds
y\;=\;-i\,\au_1\c x' \;+\;  i\, \aw_3^{}\c z'\;,
&\nonumber\\
&\ds
z\;=\;i\,\naw_1\c x' \;-\; i\, \naw_2^{-1}\c y'\;.&\label{rhs}
\end{eqnarray}
Let now rhs (\ref{rhs}) be equivalent to lhs (\ref{lhs}).
This gives eight equations for the parameters of $V_k$ and
$\ac V_k$. The solution is following: introduce
\begin{eqnarray}
\ds
\Lambda_1 &=&
u_1^{-1}\c w_1^{}\;+\;
\u_3^{\,-1}\c \u_1^{}\;+\;
\u_2^{}\c w_1^{}\;,
\nonumber\\
\ds
\Lambda_2 &=&
\u_2^{}\c w_2^{-1}\;+\;
w_3^{-1}\c u_2^{-1}\;+\;
u_1^{-1}\c w_2^{-1}\;
\nonumber\\
\ds
\Lambda_3 &=&
w_3^{-1}\c u_3^{}\;+\;
\w_2^{}\c u_3^{}\;+\;
\w_1^{\,-1}\c \w_3^{}\;,
\end{eqnarray}
and let $\Omega_i$, defined up to an ambiguity
$\Omega_i\rightarrow\acute\omega\c\Omega_i$, obey
\begin{eqnarray}
\ds
\Omega_1\c [\,u_2^{-1}\c\Lambda_2^{-1}\c u_1^{-1}\,]
&=&
\Omega_2\c [\,\u_1^{ }\c\Lambda_1^{-1}\c \u_2^{ }\,]\;,
\nonumber\\
\ds
\Omega_2\c [\,\w_3^{ }\c\Lambda_3^{-1}\c\w_2^{  }\,]
&=&
\Omega_3\c [\,w_2^{-1}\c\Lambda_2^{-1}\c w_3^{-1}\,]\;,
\nonumber\\
\ds
\Omega_3\c [\,w_1^{ }\c\Lambda_1^{-1}\c\u_3^{\,-1}\,]
&=&
\Omega_1\c [\,u_3^{ }\c\Lambda_3^{-1}\c\w_1^{\,-1}\,]\;,
\end{eqnarray}
where any of this equations is the rather nontrivial
sequence of two other, then
\begin{eqnarray}
&\ds
\aw_1^{}\;=\;w_2^{}\c \Omega_3^{-1}
\;,\;\;\;
\au_1^{}\;=\;\Lambda_2^{-1}\c w_3^{-1}
\;,&\nonumber\\
&\ds
\naw_1^{}\;=\;\Lambda_3^{-1}\c \w_2^{}
\;,\;\;\;
\nau_1^{}\;=\;\w_3^{\,-1}\c \Omega_2^{-1}
\;,&\label{one}
\end{eqnarray}
\begin{eqnarray}
&\ds
\aw_2^{}\;=\;\Omega_3^{}\c w_1^{}
\;,\;\;\;
\au_2^{}\;=\; \Omega_1^{}\c u_3^{}
\;,&\nonumber\\
&\ds
\naw_2^{}\;=\;\w_1^{}\c \Lambda_3^{}
\;,\;\;\;
\nau_2^{}\;=\;\u_3^{}\c \Lambda_1^{}
\;,&\label{two}
\end{eqnarray}
\begin{eqnarray}
&\ds
\aw_3^{}\;=\;\Lambda_2^{-1}\c u_1^{-1}
\;,\;\;\;
\au_3^{}\;=\;u_2^{}\c \Omega_1^{-1}
\;,&\nonumber\\
&\ds
\naw_3^{}\;=\;\u_1^{\,-1}\c\Omega_2^{-1}
\;,\;\;\;
\nau_3^{}\;=\;\Lambda_1^{-1}\c\u_2^{}
\;.&\label{three}
\end{eqnarray}
This map has the gauge degrees of the freedom, one degree in lhs and one
in rhs. Namely, the system is unchanged if one change lsh as follows
\begin{eqnarray}
&\ds
u_1^{}\rightarrow u_1^{}\c\omega^{-1}\;,\;\;\;\;
\w_1^{}\rightarrow \w_1^{}\c\omega^{-1}\;,
&\nonumber\\
&\ds
\u_2^{}\rightarrow \omega\c\u_2^{}\;,\;\;\;\;
\w_2^{}\rightarrow \omega\c\w_2^{}\;,
&\nonumber\\
&\ds
\u_3^{}\rightarrow \u_3^{}\c\omega^{-1}\;,\;\;\;\;
w_3^{}\rightarrow w_3^{}\c\omega^{-1}\;,
&
\end{eqnarray}
and rhs (this is the above mentioned shift of $\Omega$-s)
\begin{eqnarray}
&\ds
\nau_1^{}\rightarrow \nau_1^{}\c\acute\omega^{-1}\;,\;\;\;\;
\aw_1^{}\rightarrow \aw_1^{}\c\acute\omega^{-1}\;,
&\nonumber\\
&\ds
\au_2^{}\rightarrow \acute\omega\c\au_2^{}\;,\;\;\;\;
\aw_2^{}\rightarrow \acute\omega\c\aw_2^{}\;,
&\nonumber\\
&\ds
\au_3^{}\rightarrow \au_3^{}\c\acute\omega^{-1}\;,\;\;\;\;
\naw_3^{}\rightarrow \naw_3^{}\c\acute\omega^{-1}\;.
&\label{themap}
\end{eqnarray}
Thus we obtain the map $R(\ac\omega,\omega)$,
\begin{equation}
R_{1,2,3}(\ac\omega,\omega)\;:\;
\{V_1(\omega),V_2(\omega),V_3(\omega)\}\rightarrow
\{\ac V_1(\ac\omega),\ac V_2(\ac\omega),\ac V_3(\ac\omega)\}\;,
\end{equation}
where $\{\ac V_k\}$ are given by (\ref{themap}) and the gauge
ambiguity is expressed via the dependence
of $R$ on $\omega,\ac\omega$.
This $R$ gives the correspondence between two ``one dimensional''
orbits in the spaces of in- and out state spaces. A fixing of the gauge
means a rule comparing points on these orbits. Thus
in FTE there exist three-parameters in-orbit and three-parameters
out-orbit. Even if the points on in- and -out orbits of the
quadrilateral are fixed, there still are one-parameter freedoms
in left and right hand sides of FTE.

\section{Partial cases}

Mention now a possible algebraization of the system
(\ref{one}--\ref{three}). Suppose the parameters of $V_{12}$, $V_{13}$ and
$V_{23}$ commute. Demand that the parameters of
$\ac V_{12}$, $\ac V_{13}$, $\ac V_{23}$ also commute. This
immediately gives
\begin{equation}
\w\c w\;=\;w\c\w\;=\;k^2\;,\;\;\;\;
\u\c u\;=\;u\c\u\;=\;q^{-1}k^2\;,
\end{equation}
where $k^2$ is a center, and
\begin{equation}
u\c w\;=\;q\; w\c u\;.
\end{equation}
Expressions for $\Omega$-s are
\begin{equation}
\ds
\Omega_1\;=\;f^{-2}{k_2^2\over k_3^2}\c\Lambda_1^{-1}
\;,\;\;\;
\Omega_2\;=\;f^{-2}{1\over k_1^2k_3^2}\c\Lambda_2^{-1}
\;,\;\;\;
\Omega_3\;=\;f^{-2}{k_2^2\over k_1^2}\c\Lambda_3^{-1}
\;,
\end{equation}
where $f$ is also a center.
When $k$-s conserve, i. e. $f=1$, the map (\ref{one}--\ref{three})
can be realized by the known complete
operator $R$ -- matrix \cite{sbm-qd,ms-modified}.

To make it clear, put $q=1$, i. e. $u,w,\u,\w$ are numbers.
Then change
\begin{equation}
\ds u=>ku\,,\;\;\; w=>kw\,,\;\;\; \u=>k/u\,,\;\;\; \w=>k/w\,,
\end{equation}
Hence
\begin{equation}
\ds
\ac k_1\;=\;k_1f\,,\;\;\;\;
\ac k_2\;=\;{k_2\over f}\,,\;\;\;
\ac k_3\;=k_3f\,,
\end{equation}
and
\begin{eqnarray}
&\ds
\ac w_1\;=\;{f\over k_2}\;
{k_3\,w_1w_2+k_1\,w_2u_3+k_1k_2k_3\,u_3w_3\over w_3}\;,
&\nonumber\\
&\ds
\ac u_1\;=\;{k_2\over f}\;
{u_1u_2w_2\over k_1\,u_1w_2+k_3\,u_2w_3+k_1k_2k_3\,u_1w_3}\;,
&\nonumber\\
&\ds
\ac w_2\;=\;{k_2\over f}\;
{w_1w_2w_3\over k_3\,w_1w_2+k_1\,w_2u_3+k_1k_2k_3\,u_3w_3}\;,
&\nonumber\\
&\ds
\ac u_2\;=\;{k_2\over f}\;
{u_1u_2u_3\over k_3\,w_1u_2+k_1\,u_2u_3+k_1k_2k_3\,u_1w_1}\;,
&\nonumber\\
&\ds
\ac w_3\;=\;{k_2\over f}\;
{u_2w_2w_3\over k_1\,u_1w_2+k_3\,u_2w_3+k_1k_2k_3\,u_1w_3}\;,
&\nonumber\\
&\ds
\ac u_3\;=\;{f\over k_2}\;
{k_3\,w_1u_2+k_1\,u_2u_3+k_1k_2k_3\,u_1w_1\over u_1}\;.
&\label{old}
\end{eqnarray}
If we choose $f=1$, so that it is possible to put $k=1$,
then the map (\ref{old}) is explicitly the
complete functional map
of ZBB complex
(see \cite{ms-modified} for a table of two -- parameters functional
maps, all the examples there are just several specifications
of eq. (\ref{old}) with respect to $k$-s, and the case
$(iv)$ there is explicitly (\ref{old}).)

FTE is the sequence of FTE just for $k$-s.
Another possibility of choosing $f$ is the situation when $u_i=w_i=1$
is the stationary point of the map (\ref{old}),
this gives the electric network form of $f$:
\begin{equation}
f\;=\;{k_2\over k_1+k_3+k_1k_2k_3}\;.
\end{equation}
One more possibility is to choose $f$ as it is for the Onsager's model.

\section{A ``fusion''}

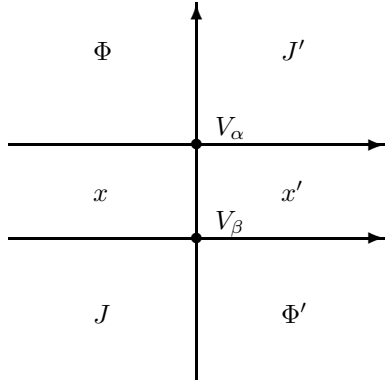
\begin{figure}
\begin{center}
\setlength{\unitlength}{0.25mm} 
\thicklines
\begin{picture}(400,200)
\put(100,0)
{  
   \begin{picture}(200,200)
   \put(100,0){\vector(0,1){200}}
   \put(0,75){\vector(1,0){200}}
   \put(0,125){\vector(1,0){200}}
   \put(100,75){\circle*{5}}
   \put(100,125){\circle*{5}}
   \put(45,30){$J$}\put(45,170){$\Phi$}
   \put(145,30){$\Phi'$}\put(145,170){$J'$}
   \put(45,95){$x$}\put(145,95){$x'$}
   \put(110,130){$V_\alpha$}\put(110,80){$V_\beta$}
   \end{picture}
} 
\end{picture}
\end{center}
\caption{Vertex $V_{<\alpha,\beta>}$, formed by two
usual vertices. The currents $x$ and $x'$ become edge indices.}
\label{fig-LL}
\end{figure}

\begin{figure}
\begin{center}
\setlength{\unitlength}{0.25mm} 
\thicklines
\begin{picture}(450,200)
\put(00,0)
{  
   \begin{picture}(200,200)
   \put(25,0){\vector(1,2){100}}
   \put(175,0){\vector(-1,2){100}}
   \put(0,25){\vector(1,0){200}}
   \put(0,75){\vector(1,0){200}}
   \put(100,150){\circle*{5}}\put(110,145){$1$}
   \put(37.5,25){\circle*{5}}\put(27.5,30){$5$}
   \put(62.5,75){\circle*{5}}\put(52.5,80){$3$}
   \put(137.5,75){\circle*{5}}\put(142.5,80){$2$}
   \put(162.5,25){\circle*{5}}\put(167.5,30){$4$}
   \end{picture}
} 
\put(250,0)
{  
   \begin{picture}(200,200)
   \put(75,0){\vector(1,2){100}}
   \put(125,0){\vector(-1,2){100}}
   \put(0,125){\vector(1,0){200}}
   \put(0,175){\vector(1,0){200}}
   \put(100,50){\circle*{5}}\put(110,45){$1$}
   \put(37.5,175){\circle*{5}}\put(17.5,180){$3$}
   \put(62.5,125){\circle*{5}}\put(42.5,130){$5$}
   \put(137.5,125){\circle*{5}}\put(152.5,130){$4$}
   \put(162.5,175){\circle*{5}}\put(177.5,180){$2$}
   \end{picture}
} 
\put(220,100){$=$}
\end{picture}
\end{center}
\caption{Intertwining relation for $V_{<2,4>}$ and $V_{<3,5>}$.}
\label{fig-double}
\end{figure}
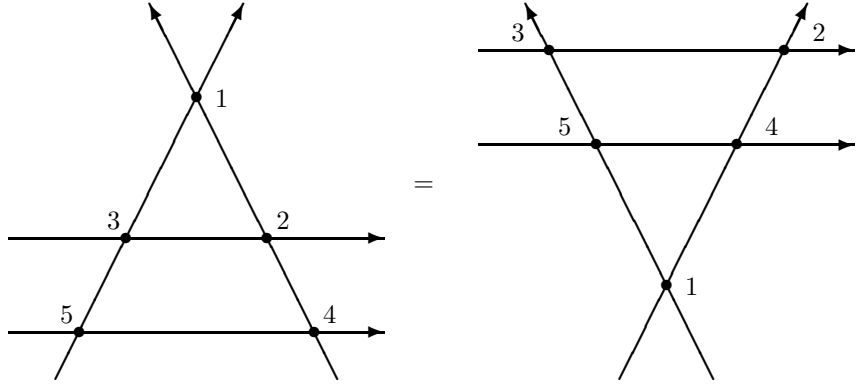

Consider a double vertex formed by two vertices,
$V_\alpha$ and $V_\beta$ as it
is shown in fig. (\ref{fig-LL}). Denote this object as
$V_{<\alpha,\beta>}$. Let the outer currents be
$J,J',\Phi,\Phi'$ and $x,x'$. They obey four relations,
\begin{eqnarray}
&\ds
J'\;=\;i\,\u_\alpha^{\,-1}\,\Phi\;,\;\;\;\;
\Phi'\;=\;-i\,u_\beta^{}\,J\;,
&\nonumber\\
&\ds
x\;=\;i\w_\beta^{}\,J-i\w_\alpha^{\,-1}\,\Phi\;,
\;\;\;
x'\;=\;-w_\beta^{-1}u_\beta^{}\,J-w_\alpha^{}\u_\alpha^{\,-1}\,\Phi\;,
&
\end{eqnarray}
so that only two of the currents are independent. Let them be $J$ and $x$,
then
\begin{eqnarray}
&\ds
x'\;=\;-(\u_\beta^{-1}+u_\alpha^{})\w_\beta^{}\,J-i u_\alpha^{}\,x\;,
&\nonumber\\
&\ds
J'\;=\;i\u_\alpha^{\,-1}\w_\alpha^{}\w_\beta^{}\,J-\u_\alpha^{\,-1}
\w_\alpha^{}\,x\;,
&\nonumber\\
&\ds
\Phi\;=\;\w_\alpha^{}\w_\beta^{}\,J+i\w_\alpha^{}\,x\;,\;\;\;
\Phi'\;=\;-iu_\beta^{}\,J\;.
&
\end{eqnarray}
Consider the transformation of a pair of such double vertices,
$V_{<2,4>}$ and $V_{<3,5>}$,
as it is shown in fig. (\ref{fig-double}). In terms of $R$
operators, this transformation is given by
$R_{123}\c R_{145}$. One can impose an invariant condition
for left and right hand sides of (\ref{fig-double}), namely,
consider the condition when in $V_{<\alpha,\beta>}$
the "edge" currents, $x$ and $x'$, are proportional, i. e.
$\u_\beta^{\,-1}+u_\alpha^{}=0$, so that $x'=-iu_\alpha^{} x$.
Obviously, if the "edge" current flows through the left hand side
of (\ref{fig-double}), then it has to flow through the right hand side
of (\ref{fig-double}). This means that if one imposes in the left hand
side the conditions
\begin{equation}\label{xx}
u_2^{}+\u_4^{\,-1}\;=\;0\;,\;\;\;\;
u_3^{}+\u_5^{\,-1}\;=\;0\;,
\end{equation}
then one obtains
\begin{equation}\label{yy}
\au_2^{}+\nau_4^{\,-1}\;=\;0\;,\;\;\;\;
\au_3^{}+\nau_5^{\,-1}\;=\;0\;,
\end{equation}
i. e. (\ref{xx}) is the ideal of $R_{123}\c R_{145}$.
For this simple system this fact can be easily verified directly.

\begin{figure}
\begin{center}
\setlength{\unitlength}{0.25mm} 
\thicklines
\begin{picture}(400,200)
\put(100,0)
{  
   \begin{picture}(200,200)
   \put(25,75){\vector(1,0){150}}
   \put(25,125){\vector(1,0){150}}
   \put(75,25){\vector(0,1){150}}
   \put(125,25){\vector(0,1){150}}
   \put(75,75){\circle*{5}}
   \put(75,125){\circle*{5}}
   \put(125,75){\circle*{5}}
   \put(125,125){\circle*{5}}
   \put(55,135){$\beta$}
   \put(55,55){$\alpha$}
   \put(135,55){$\gamma$}
   \put(135,135){$\delta$}
   \put(5,95){$x$}
   \put(95,5){$y$}
   \put(175,95){$x'$}
   \put(95,175){$y'$}
   \put(160,160){$J'$}
   \put(30,30){$J$}
   \put(30,160){$\Phi$}
   \put(155,30){$\Phi'$}
   \end{picture}
} 
\end{picture}
\end{center}
\caption{ The quadrat $V_{<\alpha,\beta,\gamma,\delta>}$.}
\label{fig-quad}
\end{figure}
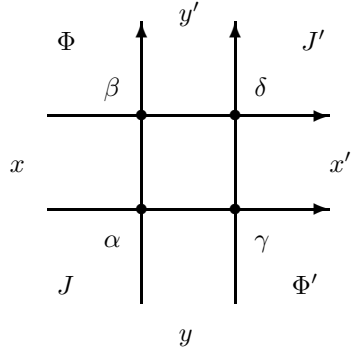

Consider now more complicated case of the quadrat (see fig.
(\ref{fig-quad}). Call this object $V_{<\alpha,\beta,\gamma,\delta>}$.
In this case three currents are independent, choose
them be $J$, $x$ and $y$. Solving the system
\begin{eqnarray}
&\ds
x\;=\;i\,\w_\alpha^{}\,J\,-\,i\,\w_\beta^{\,-1}\Phi\;,\;\;\;
x'\;=\;i\,w_\gamma^{}\,J'\,-\,i\,w_\delta^{-1}\,\Phi'\;,
&\nonumber\\
&\ds
y\;=\;i\,u_\delta^{-1}\,\Phi'\,-\,i\,u_\alpha^{}\,J\;,\;\;\;
y'\;=\;i\,\u_\beta^{\,-1}\,\Phi\,-\,i\,\u_\gamma^{}\,J'\;,
&\nonumber\\
&\ds
w_\alpha^{-1}u_\alpha^{}\,J+w_\beta^{}\u_\beta^{\,-1}\,\Phi+
u_\gamma^{-1}w_\gamma^{}\,J'+\u_\delta^{}w_\delta^{-1}\,\Phi'\;=\;0\;,
&
\end{eqnarray}
we obtain
\begin{eqnarray}
&\ds
J'\;=\;-w_\gamma^{-1}u_\gamma^{}\,(\chi\,J
\,+\,i\,u_\beta^{}\,x\,-\,i\,\w_\delta^{}\,y)\;,
&\nonumber\\
&\ds
\Phi\;=\;\w_\beta^{}\,\w_\alpha^{}\,J\,+\,i\,\w_\beta^{}\,x\;,\;\;\;
\Phi'\;=\;u_\delta^{}\,u_\alpha^{}\,J\,-\,i\,u_\delta^{}\,y\;,
&\nonumber\\
&\ds
x'\;=\;-\,i\,(w_\delta^{-1}u_\delta^{}u_\alpha^{}\,+\,u_\gamma^{}\,\chi)\,J
\,+\,u_\gamma^{}\,u_\beta^{}\,x
\,-\,(u_\gamma\,+\,\u_\delta^{\,-1})\,\w_\delta^{}\,y\;,
&\nonumber\\
&\ds
y'\;=\;i\,(\u_\beta^{\,-1}\w_\beta^{}\w_\alpha^{}\,+\,\w_\gamma^{}\,\chi)\,J
\,-\,(w_\beta^{-1}\,+\,\w_\gamma^{})\,u_\beta^{}\,x\,+\,
\w_\gamma^{}\w_\delta^{}\,y\;,
&
\end{eqnarray}
where
\begin{equation}
\chi\;=\;w_\alpha^{-1}u_\alpha^{}\,+\,
u_\beta^{}\w_\alpha^{}\,+\,
\w_\delta^{}u_\alpha^{}\;.
\end{equation}
Consider the intertwining of three copies of
$V_{<\alpha,\beta,\gamma,\delta>}$:
\begin{eqnarray}
&\ds
\Re_{1,2,3}\;:\;
V_{<\alpha_1,\beta_1,\gamma_1,\delta_1>}\,,\,
V_{<\alpha_2,\beta_2,\gamma_2,\delta_2>}\,,\,
V_{<\alpha_3,\beta_3,\gamma_3,\delta_3>}
&\nonumber\\
&\ds
\;\rightarrow\;
\ac V_{<\alpha_1,\beta_1,\gamma_1,\delta_1>}\,,\,
\ac V_{<\alpha_2,\beta_2,\gamma_2,\delta_2>}\,,\,
\ac V_{<\alpha_3,\beta_3,\gamma_3,\delta_3>}\;,&
\end{eqnarray}
so that
\begin{eqnarray}
&\ds
\Re_{1,2,3}\;=\;
R_{\alpha_1,\beta_2,\gamma_3}\c
R_{\delta_1,\gamma_2,\gamma_3}\c
R_{\beta_1,\beta_2,\beta_3}\c
R_{\gamma_1,\gamma_2,\beta_3}\c
&\nonumber\\
&\ds
R_{\alpha_1,\alpha_2,\delta_3}\c
R_{\delta_1,\delta_2,\delta_3}\c
R_{\beta_1,\alpha_2,\alpha_3}\c
R_{\gamma_1,\delta_2,\alpha_3}\;.
&
\end{eqnarray}
Imposing the condition of independence of $x',y'$ on $J$,
we obtain the ideal of the corresponding complicated $\Re$:
\begin{equation}\label{ideal}
w_\delta^{-1}u_\delta^{}u_\alpha^{}\,+\,u_\gamma^{}\,\chi\;=\;0\;,\;\;\;\;
\u_\beta^{\,-1}\w_\beta^{}\w_\alpha^{}\,+\,\w_\gamma^{}\,\chi\;=\;0\;.
\end{equation}
Next, ignoring the "edge" currents at all (i. e. putting them zeros),
so that
\begin{eqnarray}
&\ds
\Phi\;=\;\w_\beta\c\w_\alpha\c J\;=\;
\u_\beta\c \u_\gamma\c J'\;,
&\nonumber\\
&\ds
\Phi'\;=\;u_\delta\c u_\alpha\c J\;=\;
w_\delta\c w_\gamma\c J'\;,
&
\end{eqnarray}
we obtain on the surface of (\ref{ideal}) the morphism
\begin{equation}
V_\alpha\times V_\beta\times V_\gamma\times
V_\delta\Leftrightarrow V
\end{equation}
where
\begin{eqnarray}
&\ds
\overline U \;=\;i\,\u_\beta^{}\u_\gamma^{}\;,\;\;\;\;
\overline W \;=\;-i\,\w_\beta^{}\w_\alpha^{}\;,
&\nonumber\\
&\ds
U \;=\; i\,u_\delta^{}u_\alpha^{}\;,\;\;\;\;
W \;=\; -i\,w_\delta^{}w_\gamma^{}\;.
&\label{fusion}
\end{eqnarray}
Obviously, all these manipulations resemble the fusion for
the two dimensional models. For an operator formulation, when
$R$ -- operators can be expressed in terms of quantum dilogarithms
of $w_i,u_i$, the ideals (\ref{ideal}) mean the operator projectors
commuting with $\Re$.

Note that one can ignore the face currents and consider
only $x,y\rightarrow x',y'$. Thus one obtains exactly the formulation
of the free fermionic $6$-vertex type, that was considered
in \cite{oneparam}. Most complete formulation is, of course,
the formulation with the face -- edge currents.

\section{Discussion}

In this notes we have proposed some algebraical toy which can be
interpreted as a intertwining problem
for the complex of $R$ -- operators associated with ZBB statistical model.
Obvious is only one advantage of this toy: a sort of a ``fusion''.
A nonsense (or again an advantage)  of this toy is also obvious:
the gauge ambiguity. One can fix this ambiguity in different ways,
so that $3D$ $R$ -- operators gain some non -- quantized functional
part.

Nevertheless, we guess that our system would lead to something
more general then known set of the solutions of the tetrahedron equation.
A na\"{\i}ve way to obtain other $R$ -- matrices is to regard the elements
$u_k,w_k,\u_k,\w_k$ as matrices of the same structure, and hence
noncommutative for different $k$-s.
Give only one example: consider the case when
\begin{equation}
\ds u\;=\;w\;=\;\u\;=\;\w\;=\;
\left(\begin{array}{cc}
0 & k \\ 1/s & 0\end{array}\right)\;,
\end{equation}
that is the simplest generalization of the pure electric
network system so as the gauge ambiguity is canceled,
then the rather nontrivial map
\begin{eqnarray}
&\ds
k'_1\;=\;{k_2s_3\over s_1+k_2+s_3}\,,\;\;\;
s'_1\;=\;k_3+s_2+{s_2k_3\over k_1}\,,
&\nonumber\\
&\ds
k'_2\;=\;k_1+k_3+{k_1k_3\over s_2}\,,\;\;\;
s'_2\;=\;{s_1s_3\over s_1+k_2+s_3}\,,
&\nonumber\\
&\ds
k'_3\;=\;{s_1k_2\over s_1+k_2+s_3}\,,\;\;\;
s'_3\;=\;k_1+s_2+{k_1s_2\over k_3}\,
&
\end{eqnarray}
is obtained.

\vspace{1cm}

\noindent
{\bf Acknowledgments}:
I should like to thank Yu. G. Stroganov, H. E. Boos,
V. V. Mangazeev, G. P. Pron'ko, F. W. Nijhoff
and especially
Rinat Kashaev and Igor Korepanov for many fruitful
discussions.

\noindent
The work was partially
supported by the grant of the
Russian Foundation for Fundamental research No 95 -- 01 -- 00249.


\end{document}